\documentclass{PoS}

\title{Many flavor approach to study the critical point in finite density QCD}

\ShortTitle{Many flavor approach to study the critical point in finite density QCD}

\author{
\speaker{Ryo Iwami}\\
Graduate School of Science and Technology, Niigata University, Niigata
950-2181, Japan\\
E-mail: \email{iwami@muse.sc.niigata-u.ac.jp}
}

\author{
Shinji Ejiri\\
Department of Physics, Niigata University, Niigata 950-2181, Japan\\
E-mail: \email{ejiri@muse.sc.niigata-u.ac.jp}
}

\author{
Norikazu Yamada\\
KEK Theory Center, Institute of Particle and Nuclear Studies, High
Energy Accelerator Research Organization (KEK), Tsukuba 305-0801,
Japan\\
School of High Energy Accelerator Science, SOKENDAI (The Graduate
University for Advanced Studies), Tsukuba 305-0801, Japan\\
E-mail: \email{norikazu.yamada@kek.jp}
}

\abstract{
We discuss the QCD critical point at finite density through the study of many flavor QCD, in which two light flavors and $N_f$ massive flavors exist. 
Performing simulations of QCD with two flavors of improved Wilson fermions, 
we calculate the probability distribution function of a generalized plaquette in many flavor QCD at finite temperature and density. 
The dynamical effects of massive flavors and the chemical potential are incorporated using the reweighting technique. 
From the shape of the distribution functions, we determine the critical surface separating the first order transition and crossover regions in the parameter space of the light and massive quark masses and the chemical potentials. 
It is found that the critical massive quark mass becomes larger as 
the chemical potential increases in the $(2+N_f)$ flavor QCD. 
The indication to the (2+1) flavor QCD is then discussed.
}

\FullConference{The 33rd International Symposium on Lattice Field Theory\\
                 14 -18 July  2015\\
                 Kobe International Conference Center, Kobe, Japan
}
\begin{document}

%%%%%%%%%%%%%%%%%%%%%%%%%%%%%%%%%%%%%%%%%%
\section{Introduction}
\label{introduction}

\begin{figure}[tb]
\begin{minipage}{1.0\textwidth}
\vspace{-5mm}
\begin{center}
\centerline{
\includegraphics[width=55mm,height=42mm,clip]{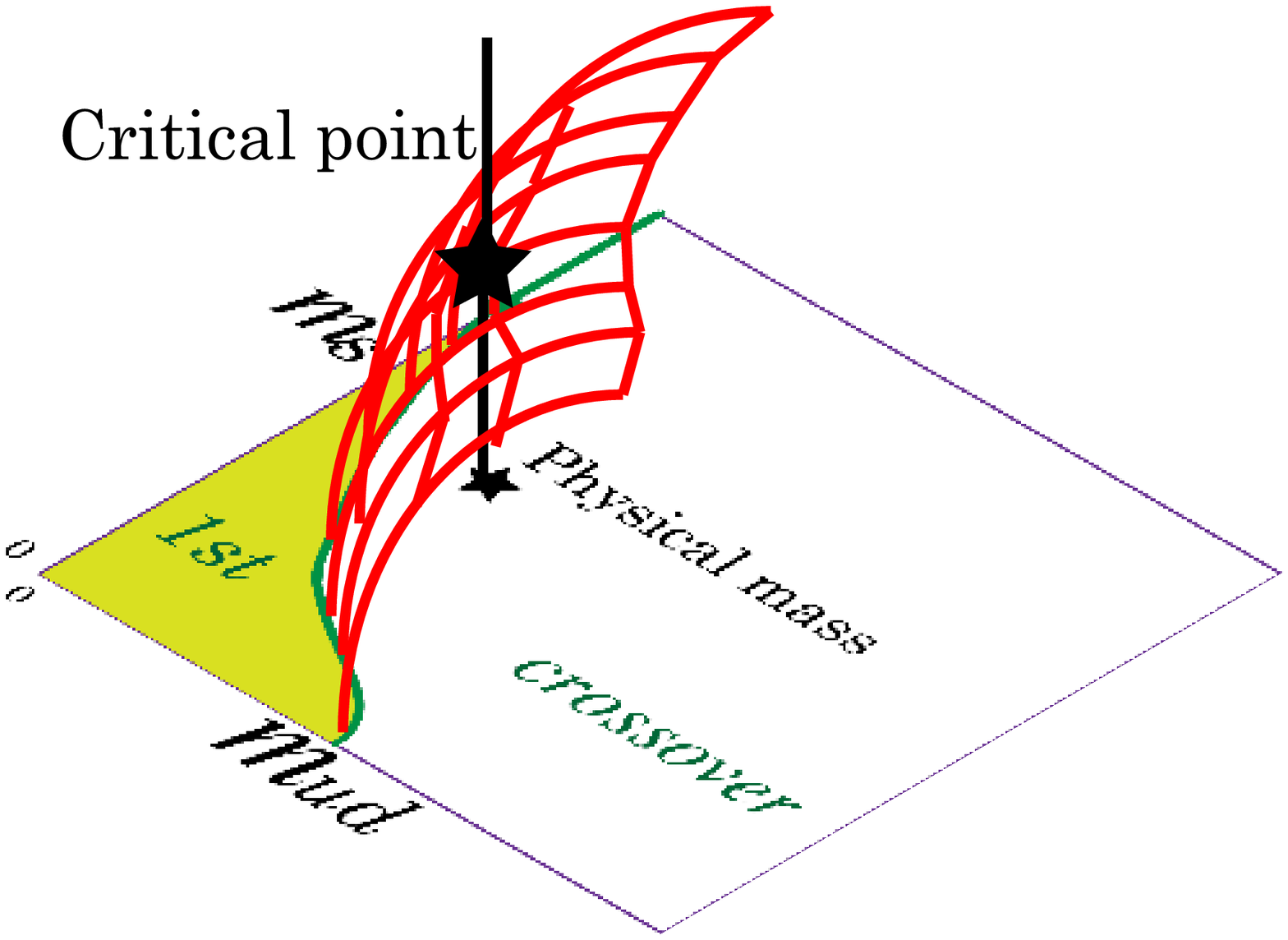}
\includegraphics[width=50mm,clip]{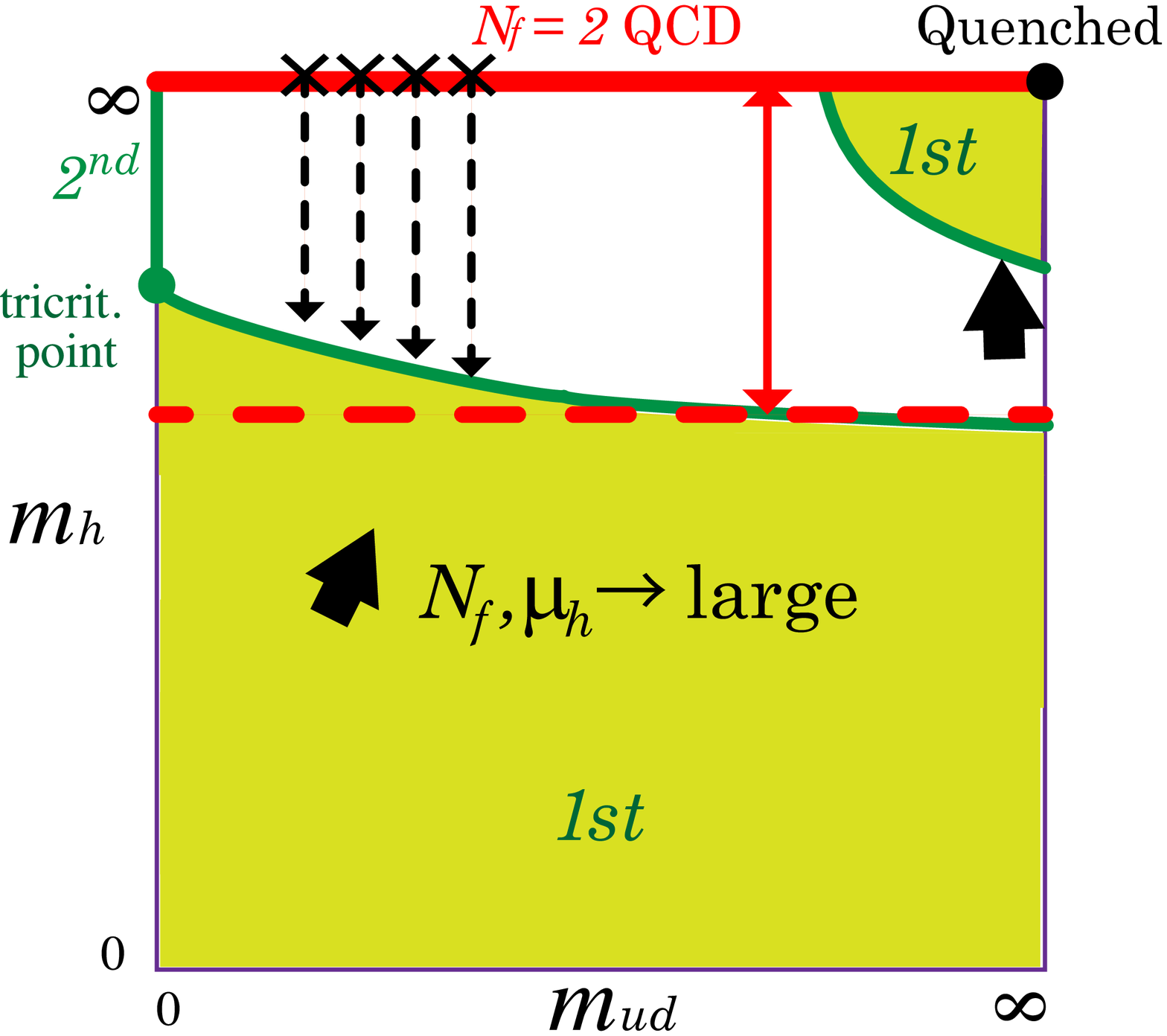}
\includegraphics[width=46mm,clip]{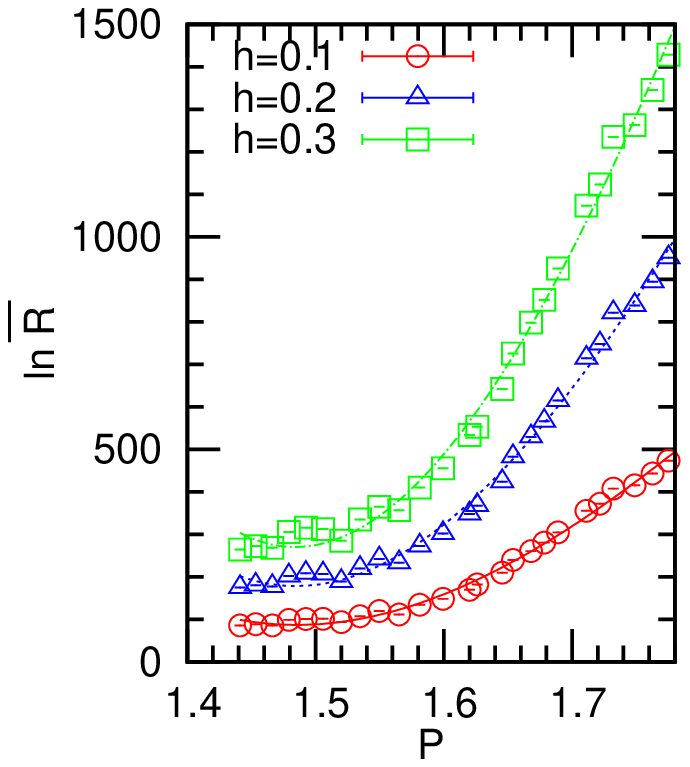}
}
%\includegraphics[width=55mm,height=42mm,clip]{./EPSPNG/columbia-2+1_3D_simple_rev20151212.eps}
%\includegraphics[width=60mm,clip]{./EPSPNG/columbia-2+1_new_phisycal_critical.eps}
%\vspace{0mm}
\caption{Left: The critical surface in the $(m_l, m_h, \mu)$ space.
Middle: Quark mass dependence of the nature of finite temperature phase transition for $(2+N_f)$ flavor QCD.
Right: $\ln \bar{R}$ as a function of $P$ for $h=0.1$ (red), $0.2$ (blue) and $0.3$ (green) at $\mu_l= \mu_h=0.$}
\label{fig:StrMany}
\end{center}
\end{minipage}
\end{figure}

One of the most interesting topics among the study of QCD is to find the critical point at finite temperature and density.
While the chiral phase transition at the physical quark masses is considered to be crossover at low density, it is expected to change into a first order transition at the critical density.
Since the study in the high density region is difficult because of the sign problem,
it will be important to investigate the boundary of the first order transition at low density in the quark mass parameter space of (2+1) flavor QCD, using the property that the nature of the transition changes also with the quark mass. 
From the information about the boundary of the first order region, it may be possible to extrapolate the boundary to the high density region for the determination of the critical point in the real world.
The standard expectation of the critical surface is illustrated in the left panel of Fig.~\ref{fig:StrMany} as a function of the light quark mass $m_l$, the strange quark mass $m_h$ and the chemical potential $\mu$. 
The critical surface is indicated by the red curves.
The light quark side is the first order region and the heavy quark side is the crossover region.
The black line represents physical quark mass.
Recent lattice QCD studies suggest that 
the first order region near the massless limit is very small at zero density.
Hence, the determination of the critical surface may be difficult.

In this paper, we study a system in which two light quarks and $N_f$ heavy quarks exist.
Performing simulations of QCD with two flavors of improved Wilson fermions, we determine the boundary of the first order region in many flavor QCD at finite temperature and density. 
The reweighting technique is used together with a hopping parameter expansion to incorporate
the dynamical effects of massive flavors and the chemical potential. 
To use the hopping parameter expansion, the hopping parameter, i.e. the inverse quark mass, must be small. 
%
%In the mass parameter space of (2+1) flavor QCD, where $m_l$ and $m_h$ are the up-down-quark mass and the strange quark mass, respectively, interesting topics are only two in the region where the strange quark is very heavy. 
%One is the first order transition when the up-down quarks are also heavy \cite{1,2}.
%The other is the finite temperature phase transition of the massless two flavor QCD.
%To understand the nature of the massless limit is a long standing problem, which is discussed in Ref.~\cite{yamada13} using our approach.
%
In Ref.~\cite{yamada13}, the critical heavy quark mass is found to increase as $N_f$ increases and the endpoint of the first order transition can be investigated in the heavy quark region, 
where the hopping parameter expansion works well, for large $N_f$.
We illustrate the quark mass dependence of the nature of phase transitions in Fig.~\ref{fig:StrMany} (middle) for (2+$N_f$) flavor QCD, where $m_l$ and $m_h$ are the masses of light flavors and $N_f$ flavors.
The yellow regions are the first order regions and the green curves are the second order critical lines. 
Moreover, we will show in this paper that the critical $m_h$ increases as $\mu$ increases. 
This suggests that we may deal with the critical curve even for (2+1) flavor in the case that the critical mass is heavy at large $\mu$.
Through the study of $(2+N_f)$ flavor QCD, we discuss the QCD critical point at finite density.

In the next section, we explain a histogram method to identify the nature of the phase transitions.
We then show our numerical results about the boundary of the first order phase transition at finite density in Sec.~\ref{sec:results}.
The conclusions are given in Sec.~\ref{sec:conclusion}.
 
%%%%%%%%%%%%%%%%%%%%%%%%%%%%%%%%%%%%%%%%%%
\section{Histogram method}
\label{sec:method}

We study the system where 
two light flavors and $N_f$ degenerate heavy flavors.
The hopping parameter and the quark chemical potential are $\kappa_l$, $\mu_l$ for the light flavor and $\kappa_h$, $\mu_h$ for the heavy flavor, respectively. 
To investigate the nature of phase transitions, we consider the probability distribution function of the average plaquette, 
\begin{eqnarray}
 w(P;\beta, \kappa_{l}, \mu_l, \kappa_h, \mu_h) 
% &=& \int \mathcal{D} U \mathcal{D} \psi \mathcal{D} \bar{\psi} 
% \delta \left( P - \hat{P} \right) e^{-S_q -S_g} \nonumber \\
 &=& \int \mathcal{D} U \delta \left( P - \hat{P} \right) 
e^{6 \beta N_{\rm site} \hat{P}} \left( \det M (\kappa_l,\mu_l) \right)^{2} 
\left( \det M (\kappa_h,\mu_h) \right)^{N_f} 
\end{eqnarray}
where $M$ is the quark matrix, 
$N_{\rm site} \equiv N^{3}_s \times N_t$ is the number of sites,
and $\beta = 6/ g^{2}_{0}$ is the coupling constant. 
$\hat{P}$ is defined from the gauge action $S_g$ as $\hat{P} = -S_{g} /(6N_{\rm site} \beta)$ and 
is called the generalized plaquette.
$\delta (P - \hat{P})$ is the delta function, which constrains the operator $\hat{P}$ to $P$.
We use the delta function approximated by
$\delta(x) \approx 1/(\Delta \sqrt{\pi})$ $\exp[-(x/\Delta)^2]$, 
where $\Delta = 0.00025$ is adopted in this study.
For convenience, we define the effective potential as 
$V_{\rm eff} (P) \equiv - \ln w(P)$.
It is rewritten as 
\begin{eqnarray}
 V_{\rm eff}(P; \beta, \kappa_{l}, \mu_l, \kappa_h, \mu_h) 
 &=& V_{0} (P;\beta_{0}, \kappa_l) - \ln R(P; \beta, \beta_0, \kappa_l, \mu_l, \kappa_h, \mu_h) 
\end{eqnarray}
with the potential of two flavor at $\mu=0$, $V_0 (P;\beta_0, \kappa_l)$, and
\begin{eqnarray}
 \ln R (P) 
 &=& 6 (\beta - \beta_0) N_{\rm site} P + \ln \left\langle \left( \frac{ \det M (\kappa_l, \mu_l) }{ \det M (\kappa_l, 0) } \right)^{2} \left(    \frac{ \det M (\kappa_h, \mu_h) }{ \det M (0, 0) } \right)^{N_f} \right\rangle_{(P; {\rm fixed})} ,
\label{eq:lnr}
\end{eqnarray}
where $ \left\langle \cdots \right\rangle_{(P; {\rm fixed})}
= \left\langle \delta (P - \hat{P}) \cdots \right\rangle_{(\beta_0,\kappa_l)} / \left\langle \delta (P - \hat{P}) \right\rangle_{(\beta_0, \kappa_l)} $. 
$\langle \cdots \rangle_{(\beta_0, \kappa_l)}$ means the ensemble average over
two flavor configurations generated at $\beta_0$, $\kappa_l$ and
vanishing $\mu_l$. $\beta$ may differ from $\beta_0$.
%Because the factors $e^{6 N_{site} \hat{P}}$ in the numerator and denominator cancel, 
%this $R(P)$ is independent of $\beta$.

We evaluate the quark determinant of $N_f$ flavors in $R(P)$ with the leading order of the hopping parameter expansion for the standard Wilson action \cite{whot13}. 
\begin{eqnarray}
\ln \left[ \frac{ \det M(\kappa_h, \mu_h) }{ \det M(0,0) } \right]
&=& 288 N_{\rm site} \kappa^{4}_{h} \hat{P} + 12 N^{3}_{s} ( 2 \kappa_{h} )^{N_t} \left( \cosh (\mu_h /T) \hat{\Omega}_{R} + i \sinh (\mu_h/T) \hat{\Omega}_{I} \right) + \cdots , \ \ \
\label{eq:lnrhpe}
\end{eqnarray}
where
$\hat{\Omega}_{R}$ and $\hat{\Omega}_{I}$ are the real and imaginary parts of the Polyakov loop.
This approximation is valid when $\kappa_h$ is small or the quark is heavy.
Note that no expansion in terms of $\mu_h/T$ is performed. 
On the other hand, the quark determinant of light flavors is computed by a Taylor expansion in terms of $\mu_l$ \cite{BS02, ejiri08, whot09}, i.e.
\begin{eqnarray}
\ln \left[ \frac{ \det M(\kappa_l, \mu_l) }{ \det M(\kappa_l, 0) } \right]  
\approx \sum_{n=1}^{N_{\rm max}} \frac{1}{n!} 
\left[ \frac{\partial^n (\ln \det M(\kappa_l, \mu))}{\partial (\mu/T)^n} 
\right]_{\mu=0} \left( \frac{\mu_l}{T} \right)^n .
\label{eq:Tay}
\end{eqnarray}
Because the sign problem is serious for the calculation of $R(P)$ at high density, 
the method discussed in Ref.~\cite{ejiri08, whot09} is used to avoid the sign problem,
\begin{eqnarray}
R(P) = \left\langle e^{\hat{F}} \right\rangle_{(P: {\rm fixed})} \left\langle e^{i \hat{\theta}} 
\right\rangle_{(P, F: {\rm fixed})}
\approx \left\langle e^{\hat{F}} \right\rangle_{(P: {\rm fixed})}
\exp \left( -\frac{1}{2} \langle \hat{\theta}^2 \rangle_{(P: {\rm fixed})} \right),
\label{eq:rcum}
\end{eqnarray}
%
%\begin{eqnarray}
%R(P) &=&
%\left\langle \left| \frac{\det M(\mu_l)}{\det M(0)} \right|^2 
%\left| \frac{\det M(\mu_h)}{\det M(0)} \right|^{N_{\rm f}}
%\right\rangle_{(P: {\rm fixed})} \left\langle e^{i \hat{\theta}} 
%\right\rangle_{(P, F: {\rm fixed})}
%\nonumber \\
%& \approx & \left\langle \left| \frac{\det M(\mu)}{\det M(0)} \right|^2 
%\left| \frac{\det M(\mu)}{\det M(0)} \right|^{N_{\rm f}}
%\right\rangle_{(P: {\rm fixed})}
%\exp \left( -\frac{1}{2} \langle \hat{\theta}^2 \rangle_{(P: {\rm fixed})} \right) ,
%\label{eq:rcum}
%\end{eqnarray}
%
where $\hat{F}= \ln \left( \left| \frac{\det M(\mu_l)}{\det M(0)} \right|^2 
\left| \frac{\det M(\mu_h)}{\det M(0)} \right|^{N_{\rm f}} \right)$ and $\hat{\theta}$ is the complex phase of 
$\left( \frac{\det M(\mu_l)}{\det M(0)} \right)^2 
\left( \frac{\det M(\mu_h)}{\det M(0)} \right)^{N_{\rm f}}$. 
In this analysis, we assume that the distribution of $\hat{\theta}$ is well approximated by a Gaussian function and the higher order cumulants can be neglected.

The first term of Eq.~(\ref{eq:lnrhpe}) that is proportional to $\hat{P}$ can be absorbed into the gauge action by shifting $\beta \to \beta^* \equiv \beta + 48 N_f \kappa^{4}_h$.
The reweighting factor is written as  
$\ln R (P; \beta, \kappa_h, \mu_h) = \ln \bar{R} (P; h, \mu_h) + {\rm (plaquette \ term)} 
+ O (\kappa^{N_t +2}_{h})$
for $\mu_l=0$, for example, with
\begin{eqnarray}
\bar{R}(P;h,\mu_h)
 = \left\langle \exp \left[ 6 h N^{3}_s \left( \hat{\Omega}_{R} 
+ i \tanh \left( \frac{\mu_h}{T} \right) \hat{\Omega_{I}} \right) \right] 
\right\rangle_{(P:{\rm fixed}, \beta_0)}
\end{eqnarray}
where $h=2N_f (2\kappa_h)^{N_t} \cosh (\mu_h /T)$.
%$\bar{R} (P;h,\mu_h)$ is independent of $\beta_0$. 
$\bar{R} (P; h, \mu_h)$  depends on the parameters only through $\beta^*$, 
$h$ and $\tanh (\mu_h /T)$.
The second term is the complex phase of the reweighting factor, which is proportional to $\tanh (\mu_h /T)$.
Because $| \tanh (\mu_h /T) |$ is smaller than one, the $\tanh (\mu_h /T)$ dependence can be easily controlled \cite{whot13}.

We then find the critical $h$, at which the first order transition terminates, instead of the critical $\kappa_h$.
At a first order transition point, $V_{\rm eff}$ shows 
a double-well shape as a function of $P$, 
and equivalently the curvature of the potential $d^2 V_{\rm eff}/dP^2$ 
is negative around the center of the double-well potential.
Moreover, the curvature $d^2 V_{\rm eff}/dP^2$ is independent of $\beta$, since $\beta$ appears only in the linear term of $P$ in the right hand side of Eq.~(\ref{eq:lnr}).
Although $\beta$ must be adjusted to the first order transition point to observe the double-well potential, the fine tuning is not necessary if we investigate the curvature \cite{ejiri08}.
Hence, we investigate the curvature of the potential and find the boundary of the first order phase transition. 
The curvature is given by 
\begin{eqnarray}
\frac{d^2 V_{{\rm eff}}}{dP^2}(P) 
= \frac{d^2 V_{0}}{dP^2} (P) - \frac{d \ln \bar{R}}{dP^2} (P). 
\end{eqnarray}
The curvature of the two flavor effective potential $V_0$ can be estimated from the relation between the plaquette susceptibility 
$\chi_P = 6 N_{\rm site} \langle ( \hat{P} - \langle \hat{P} \rangle )^2 \rangle$ 
and the curvature of the potential, 
\begin{eqnarray}
\frac{d^2 V_0}{dP^2} = \frac{6N_{\rm site}}{\chi_P}, 
\label{eq:curpot2}
\end{eqnarray}
assuming the Gaussian distribution. 
Because the finite temperature transition is crossover for two flavor QCD with finite quark mass, the distribution function is expected to be Gaussian.
%, i.e. the effective potential is a quadratic function. 
The second derivative $d^2 \ln R/dP^2$ is computed by fitting the data of $\ln \bar{R}$ to a sixth degree function.
After finding the critical $h$ for each $\kappa_l$, $\mu_l$ and $\tanh (\mu_h /T)$, 
we determine the critical $\kappa_h$ from the equation 
$h=2N_f (2\kappa_h)^{N_t} \cosh (\mu_h /T)$.

%%%%%%%%%%%%%%%%%%%%%%%%%%%%%%%%%%%%%%%%%%
\section{Chemical potential dependence of the critical mass in $(2+ N_f)$ flavor QCD}
\label{sec:results}

We have performed simulations of QCD with degenerate two flavor $O(a)$-improved Wilson quark and RG-improved Iwasaki gauge actions at zero density in Ref.~\cite{yamada15}.
The lattice size $N_{\rm site}$ is $16^{3} \times 4$.
We adopt four hopping parameters of light quarks $\kappa_l$ and 25 to 32 $\beta$ values at each $\kappa_l$, which cover the pseudo-critical $\beta$, and 
10,000 to 40,000 trajectories have been accumulated at each simulation point.
The details of the simulation parameter are shown in Ref.~\cite{yamada15}.
Among the configurations, the finite density analysis is carried out with $\kappa_l = 0.1450$ and $0.1475$. 
We use 500 configurations taken every 10 trajectories at each simulation point.
The light quark determinant is evaluated up to $O(\mu_l^2)$, i.e. $N_{\rm max}=2$ in Eq.~(\ref{eq:Tay}).
We have carried out the zero temperature simulations on $16^3 \times 32$ lattices 
to determine the pseudo-scalar and vector meson mass ratio $(m_{{\rm PS}} / m_{{\rm V}})$ and 
the quark mass ($m_{\rm pcac} a$) at the psuedo-critical point.
%, determined at the peak of the susceptibility for the generalized plaquette.
These two values of $\kappa_l = 0.1450, 0.1475$ correspond to 
$m_{{\rm PS}} / m_{{\rm V}}= 0.66, 0.58$ and $m_{\rm pcac} a = 0.053, 0.035$, respectively.
%In the calculation of $\bar{R} (P;h,0,0)$, we use the delta function 
%approximated by 
%$\delta (x) \simeq 1/\left(\Delta \sqrt{\pi} \right) 
%\exp \left[ - (x/\Delta)^{2}\right]$, where $\Delta = 0.00025$ is adopted.

\begin{figure}[t]
\begin{center}
\vspace*{-5mm}
\centerline{
\hspace*{-3mm}
\includegraphics[width=60mm,clip]{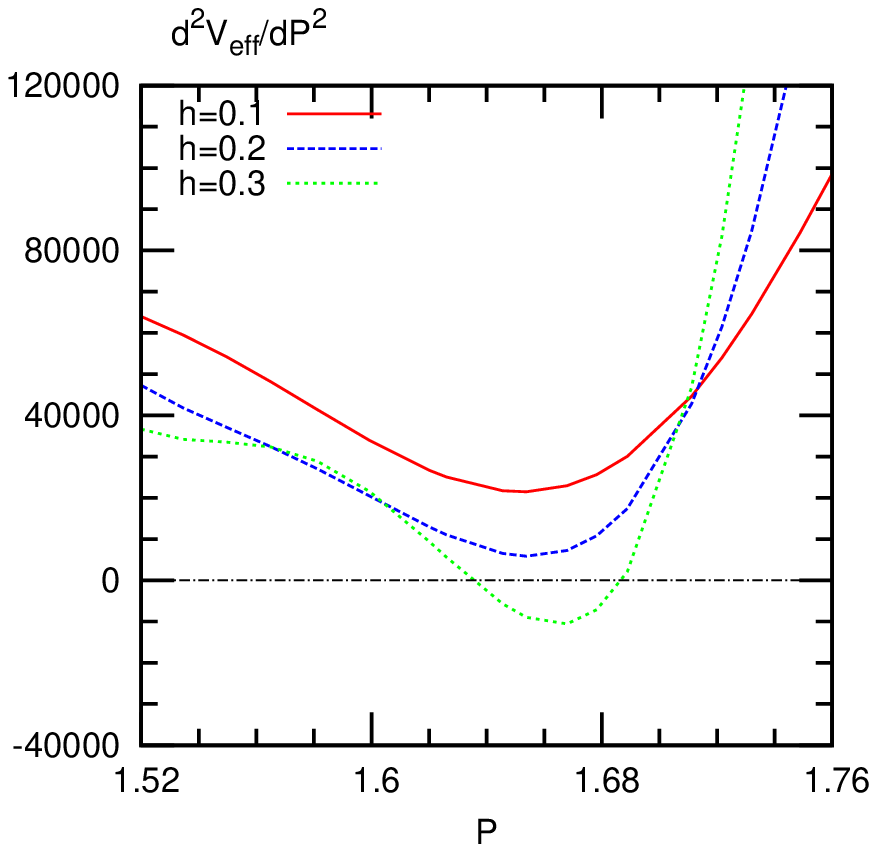}
\hspace{-6mm}
\includegraphics[width=52mm,clip]{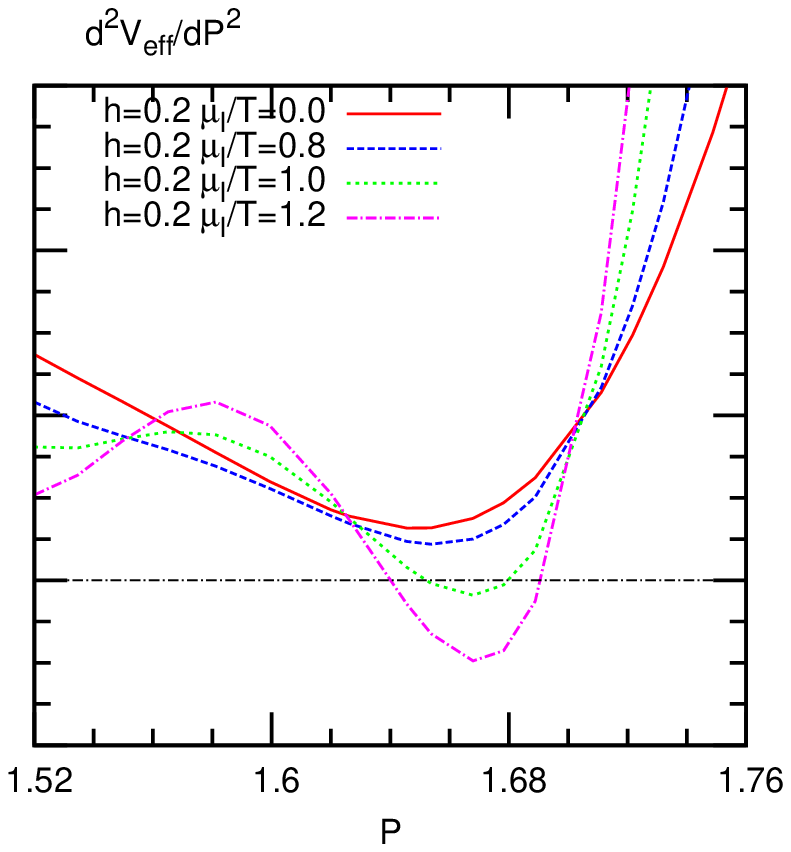}
\hspace{-6mm}
\includegraphics[width=52mm,clip]{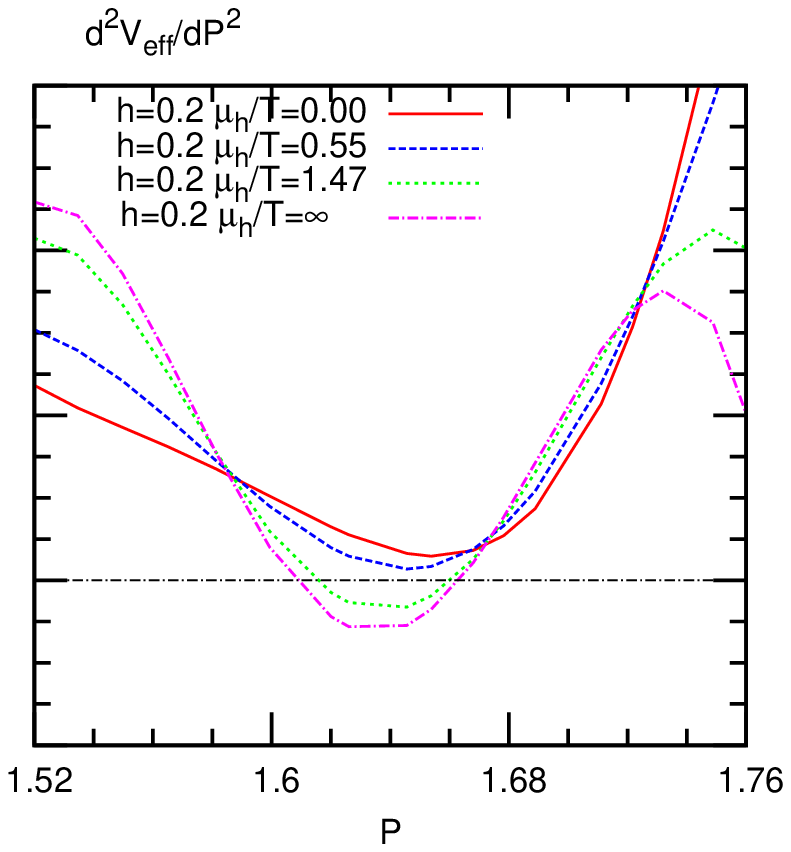}
}
%\vspace{-3mm}
\caption{The curvatures of $V_{{\rm eff}}$ for $h=0.1, 0.2, 0.3$ at $\mu_l =\mu_h =0$ (left),
for $\mu_l /T=0.0, 0.8, 1.0, 1.2$ at $h=0.2$ with $\mu_h =0$ (middle), and
for $\mu_h /T=0.0, 0.55, 1.47, \infty$ at $h=0.2$ with $\mu_l =0$ (right).}
\label{fig:curvmu}
\end{center}
\end{figure}

\begin{figure}[t]
\begin{center}
\vspace*{-5mm}
\centerline{
\includegraphics[width=80mm,clip]{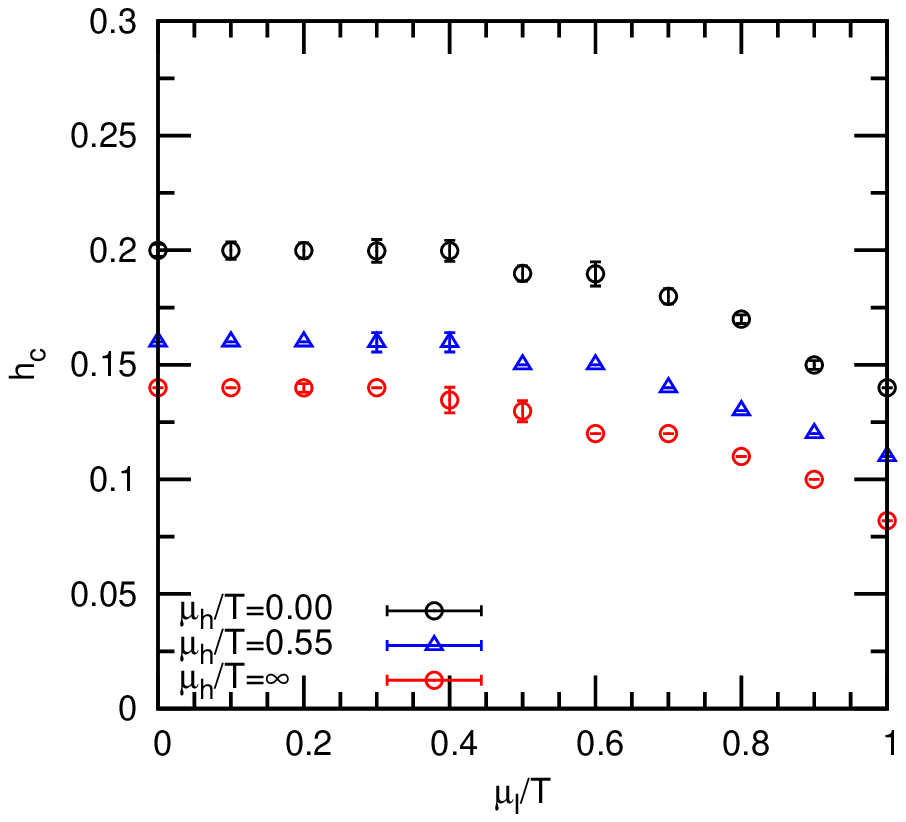}
\hspace*{-3mm}
\includegraphics[width=80mm,clip]{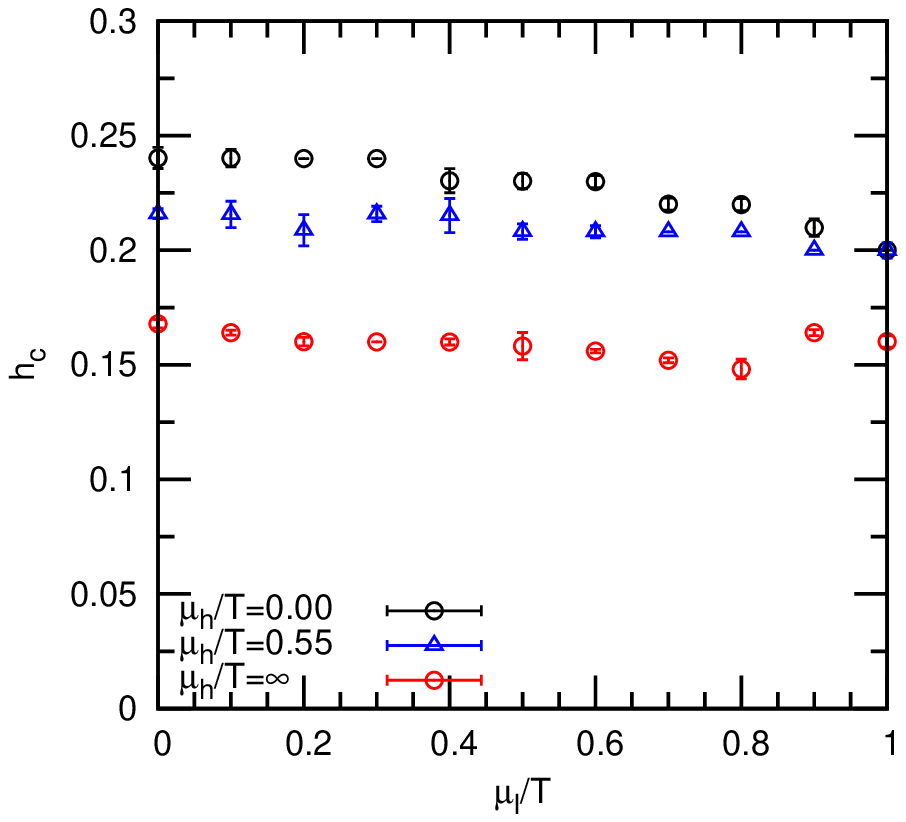}
}
\vspace{-2mm}
\caption{The critical points $h_c$ as functions of $\mu_l /T$ for $\mu_h /T = 0.00$ (black), $0.55$ (blue) and $\infty$ (red) 
at $\kappa_l$ = $0.1475$ (left) and $0.1450$ (right).}
\label{fig:critline}
\end{center}
\end{figure}

The right panel of Fig.~\ref{fig:StrMany} shows the results of $\ln \bar{R}(P)$ at $\kappa_l=0.145$ and $\mu_l=\mu_h=0$ for $h=0.1, 0.2$ and $0.3$.
$\ln \bar{R} (P)$ increases rapidly around $P \approx 1.6$ and the gradient becomes lager as $h$ increases.
The second derivatives $d^{2} \ln \bar{R} / dP^{2}$ is calculated by fitting the data of 
$\ln \bar{R}$ to a sextic polynomial of $P$ and $d^2V_0/dP^2$ is computed from $\chi_P$ 
as explained in Sec.~\ref{sec:method}.
The results of $d^2 V_{\rm eff}/dP^2$ are plotted in the left panel of Fig.~\ref{fig:curvmu} for $h=0.1, 0.2$ and $h=0.3$ at $\kappa_l=0.145$ and $\mu_l=\mu_h=0$.
This figure shows that $d^2 V_{\rm eff} / dP^{2}$ is positive for all $P$ at small $h$, while $d^2 V_{\rm eff} / dP^{2}$ becomes negative in a range of $P$ at $h=0.3$.
This indicates that the phase transition changes from crossover to first order between $h=0.2$ and $0.3$.

On the other hand, the first order transition arises also at large $\mu_l$ and $\mu_h$ even when $h$ is small.
The middle and right panels of Fig.~\ref{fig:curvmu} show the $\mu_l$ and $\mu_h$ dependence of the curvature of $V_{\rm eff}$ with $h=0.2$ and $\kappa_l=0.145$. 
The curvature becomes negative in a range of $P$ at large $\mu_l$ or $\mu_h$.
We determine the critical value of $h$, $h_c$, at which the negative curvature appears. 
The preliminary results of $h_c$ are shown in the left and right panels of Fig.~\ref{fig:critline} as functions of $\mu_l$ for $\kappa_l=0.145$ and $0.1475$, respectively. 
The shape of effective potential is double well in the first order region above $h_c$. 
The systematic error which arise in this analysis has not been estimated yet. The error bar represents the statistical error only.
%
%As we explained, the parameters which control the system are $\beta$, $\kappa_l$, $\mu_l$, $h=2N_f (2\kappa_h)^{N_t} \cosh (\mu_h /T)$ and $\tanh (\mu_h/T)$. 
%$\beta$ must be adjusted at the transition temperature to investigate the phase transition although $d^2 V_{\rm eff} / dP^{2}$ is independent of $\beta$.

Because we have used the approximation by Taylor expansion up to $O(\mu_l^2)$, the analysis is valid only in the region where $\mu_l/T$ is small.
On the other hand, the $\mu_h$ dependence is easy to investigate, since $| \tanh (\mu_h/T) |$ is smaller than one. 
The results of ``$\mu_h/T=\infty$'' mean those of $\tanh (\mu_h/T)=1$.
It is found that the first order region becomes wider as $\mu_l$ and $\mu_h$ increase.
The qualitative behavior, i.e. $h_c$ decreases as $\mu_l$ increases, is consistent with the previous results obtained by an improved staggered fermion action \cite{yamada13}.
However, the difference from the result of $h_c$ by the staggered fermion is not small quantitatively. 
Although the systematic error is not yet estimated, the discretization error of our result by the $N_t=4$ lattice may be large. 

\begin{figure}[t]
\begin{center}
\vspace*{-10mm}
\centerline{
\hspace*{-6mm}
\includegraphics[width=76mm,clip]{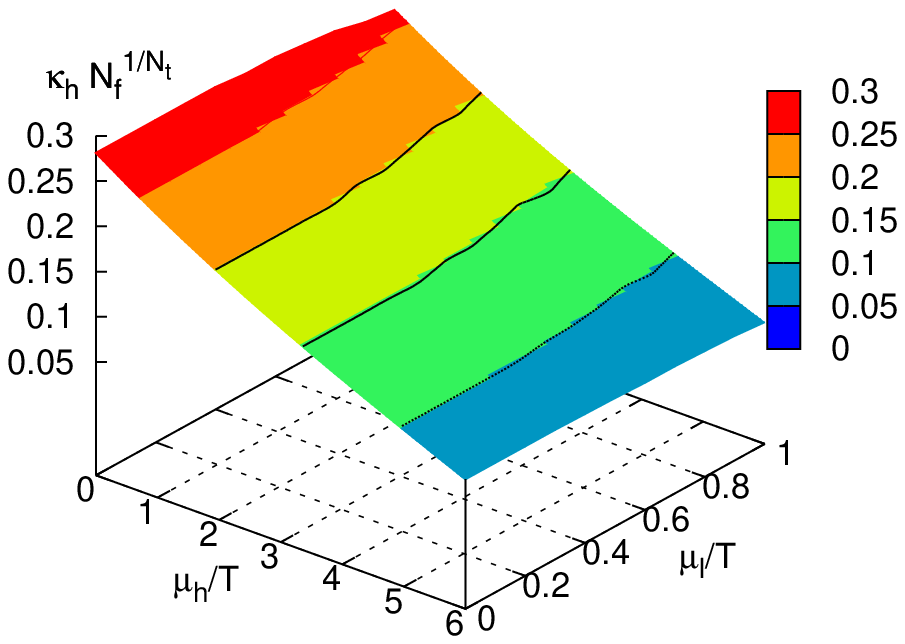}
\hspace*{-2mm}
\includegraphics[width=76mm,clip]{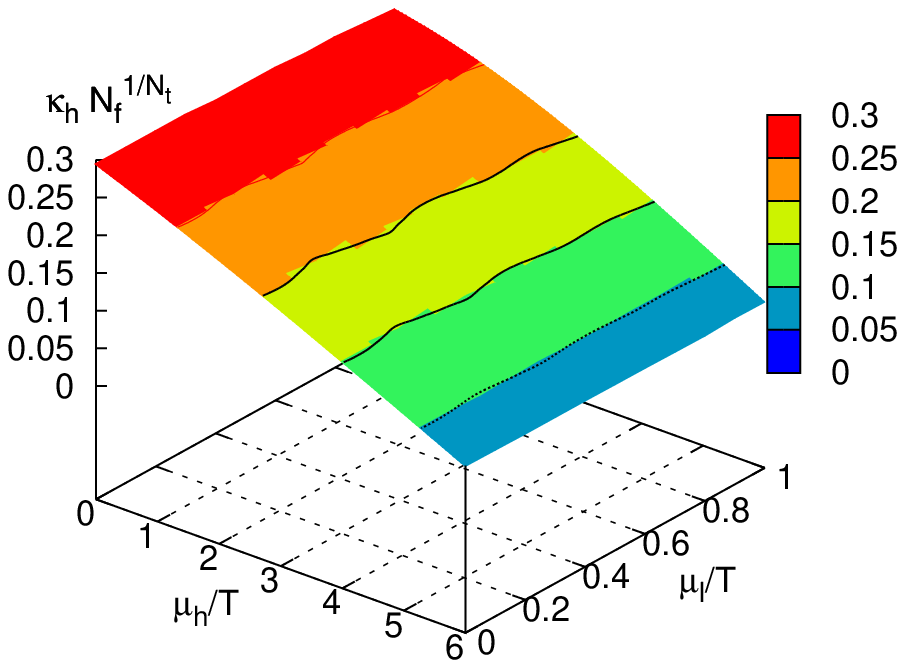}
}
\vspace{-5mm}
\caption{The critical surface in the $(\kappa^{c}_{h} N_f^{1/N_t}, \mu_l /T, \mu_h /T)$ 
space at $\kappa_l = 0.1475$ (left) and $0.1450$ (right).}
\label{fig:surface}
\end{center}
\end{figure}

Another interesting point is that the critical value $h_c$ is less dependent on $\tanh(\mu_h/T)$.
The difference between the results of $\tanh(\mu_h/T)=0$ and $1$ is only about $30\%$.
This situation is similar to the case of the critical surface in the heavy quark region \cite{whot13}.
Although the change of $h_c$ is small, the critical $\kappa_h$ changes as $\mu_h$ increases, of course. 
Rewriting $h$ as 
\begin{equation}
 \kappa_{h} N^{1/N_t}_f = \frac{1}{2} \left( \frac{h}{2 \cosh (\mu_h / T)} \right)^{1/N_t}
  ~ \to ~ \frac{1}{2} h^{1/N_t} e^{-\mu_h / (N_t T)} 
\ \ {\rm for} \ \  \left( \frac{\mu_h}{T} \gg 1 \right),
\end{equation}
one finds that the critical $\kappa_h$ decreases exponentially as $\mu_h$ for large $\mu_h$ when $\tanh(\mu_h/T)$ dependence of $h_c$ is small.
We then plot the critical value of $\kappa^{c}_{h} N_f^{1/N_t}$ in Fig.~\ref{fig:surface} as a function of $\mu_l/T$ and $\mu_h/T$ translating Fig.~\ref{fig:critline}. 

Because the approximation of $O((\mu_l/T)^2)$ is used, $\mu_l/T$ must be small in this analysis,
whereas, the critical points are determined for $0 \leq \tanh(\mu_h/T) \leq 1$, i.e. 
$0 \leq \mu_h/T \leq \infty$.
Then, the critical value of $\kappa^{c}_{h} N_f^{1/N_t}$ decreases with $\mu_h/T$ exponentially. 
This result derives an important consequence.
To apply the hopping parameter expansion, we assume $\kappa_h$ to be small. 
This condition is always valid for large $N_f$ even if the critical $h$ is large.
However, for large $\mu_h$, $N_f$ is not needed to be large because 
$\kappa^{c}_{h} N_f^{1/N_t}$ is no longer large. 
This means the analysis based on the hopping parameter expansion can be applied for small $N_f$.
If we assume that the hopping parameter expansion is valid when the critical $\kappa_{h}$ is smaller than 0.1, 
%($h_c \le 0.032 \times N_f \cosh (\mu_h/T)$),
for example, one can investigate the critical value of $\kappa_h$ at $\mu_h/T > 5.0$ in $(2+1)$ QCD, i.e. $N_f=1$.
Thus, the critical surface of $(2+1)$ flavor QCD can be studied when the density of strange quark is very large in our approach. 
This suggests that the extrapolation from the high strange quark density region may be a possible way to determine the QCD critical point.

%%%%%%%%%%%%%%%%%%%%%%%%%%%%%%%%%%%%%%%%%%
\section{Conclusions}
\label{sec:conclusion}

We studied the phase structure of the many flavor system, in which two light flavors and $N_f$ massive flavors exist, 
performing simulations of two flavor QCD with improved Wilson fermions and 
using the hopping parameter expansion of the heavy quarks. 
Through the shape of the distribution functions, we determined the critical surface separating the first order transition and crossover regions.
It is found that the critical mass of the massive quarks becomes larger as the chemical potential increases. 
Because we approximate the light quark determinant by a Taylor expansion in terms of the light quark chemical potential $\mu_l$, the application range of $\mu_l$ is limited to a small region. 
On the other hand, the heavy quark chemical potential $\mu_h$ can be controlled in a wide range, and it is found that the critical $\kappa_h$ decreases exponentially with $\mu_h$ at large $\mu_h$.
Since the critical $\kappa_h$ is sufficiently small at large $\mu_h$, the hopping parameter expansion is applicable even for $(2+1)$ flavor. 
This conclusion about the critical $\kappa_h$ at large $\mu_h$ is valid for $(2+1)$ flavor QCD.

\paragraph{Acknowledgments}
This work is in part supported by
JSPS KAKENHI Grant-inAid for Scientific Research (Nos.\ 15H03669, 26287040, 26400244) 
and by the Large Scale Simulation Program of High
Energy Accelerator Research Organization (KEK) No.\ 14/15-23.

\end{document}